

\hsize=6.5in
\vsize=9in
\tolerance 1500
\interlinepenalty=1000
\clubpenalty=1000
\widowpenalty=1000

\font\headfont=cmbx10 scaled \magstep2
\font\subfont=cmbx10 scaled \magstep1
\font\ninesc=cmcsc9

\def\efoot#1{${}^{#1}$}
\def\ie{{\sl i.e.\/}~}
\def\etc{{\sl etc.\/}}
\def\dxi{\dot{\xi}^i}
\def\dxj{\dot{\xi}^j}
\def\pdxi{\partial\skew{3}\dot{\xi}^{i\vrule height 1.5ex width0pt}}
\def\pdxq{\partial\skew{3}\dot{q}^{\vrule height 1.5ex width0pt}}

\def\vec#1{{\bf #1}}
\def\vnam{\setbox0=\hbox{$\nabla$}%
\copy0\kern-\wd0
\kern-.02em\copy0\kern-\wd0
\kern+.02em\raise.02ex\copy0\kern-\wd0
\kern+.02em\box0}

\topskip=0.35in
\nopagenumbers
\headline={\ifnum\pageno>1
{\ninesc (Constrained) Quantization Without Tears \hfil page \folio}
\else\hfil\fi}

\baselineskip 24pt plus 1pt

\pageno=0
\centerline{\headfont (Constrained) Quantization Without Tears}
\vskip .5in
\centerline{R. Jackiw}
\vskip .1in
\centerline{\sl Center for Theoretical Physics}
\centerline{\sl Laboratory for Nuclear Science}
\centerline{\sl Massachusetts Institute of Technology}
\centerline{\sl Cambridge, MA\ \ 02139}
\vfill
\centerline{\bf ABSTRACT}
\smallskip
\centerline
{An alternative to Dirac's constrained quantization procedure is explained.}
\vfill
\centerline{Presented at:}
\centerline{``Constraint Theory and Quantization Methods,''}
\centerline{Montepulciano, Italy, June 1993}
\vskip.2in
\hbox to \hsize{\hrulefill}
\medskip
\line{CTP\#2215 \hfil hep-th/9306075 \hfil May 1993}
This work is supported in part by funds provided by the
U. S. Department of Energy (D.O.E.) under contract \#DE-AC02-76ER03069.

\eject

\vskip .1in
\centerline{\subfont (Constrained) Quantization Without Tears}
\centerline{\rm R. Jackiw}
\vskip .2in
\centerline{\bf ABSTRACT}
\centerline
{An alternative to Dirac's constrained quantization procedure is explained.}

\vskip.2in

To accomplish conventional and elementary quantization of a dynamical system,
one is instructed to: begin with a Lagrangian, eliminate velocities in favor
of momenta by a Legendre transform that determines the Hamiltonian, postulate
canonical brackets among coordinates and momenta and finally define dynamics
by commutation with the Hamiltonian.  But this procedure may fail for several
reasons: it may not be possible to solve for the velocities in terms of the
momenta, or it may be that the Hamiltonian equations do not reproduce the
desired dynamical equations.  In such cases one is dealing with so-called
``singular'' Lagrangians and ``constrained'' dynamics.  Almost half a century
ago Dirac developed his method for handling this situation\efoot{1}\ and since
that time the subject has defined an area of specialization in mathematical
physics, as is put into evidence by a recent monograph\efoot{2}\ and by this
series of workshops\efoot{3}.

While Dirac's approach and subsequent developments can cope with most
models of interest, my colleague Ludwig Faddeev and I realized
that in many instances Dirac's method is unnecessarily cumbersome and can be
streamlined
and simplified.  We have advertised\efoot{4}\ an alternative approach, based on
Darboux's theorem, wherein one arrives at the desired results --- formulas for
brackets and for the Hamiltonian --- without following Dirac step by step.

Very specifically, two aspects of the Dirac procedure are avoided.  First,
when it happens that the Lagrangian $L$
depends linearly on the velocity $\dxi$
for one of the dynamical variables $\xi^i$,
or even is independent of $\dxi$, the attempt to define the canonical momentum
$\Pi_i = {\partial L\over\pdxi}$, and to eliminate
$\dot{\xi}_i$ in favor of $\Pi_i$ obviously fails.  In the Dirac procedure,
one nevertheless defines a canonical momentum and views the
$\dxi$-independent expression ${\partial L\over\pdxi}$
as a constraint on $\Pi_i$.  In our method, such constraints are never
introduced.  Second, in the Dirac procedure constraints are classified and
distinguished as first class or second class, primary or secondary.
This distinction is not made in our method; all constraints are held to the
same standard.

It is therefore clear that our approach eliminates useless paperwork, and here
I shall give a description with the hope that this audience of specialists
will appreciate the economy of our proposal and will further adopt and
disseminate it.

I shall use notation appropriate to a mechanical system, with coordinates
labeled by $\{i, j, \ldots \}$ taking values in a set of integers up to $N$,
and a summation convention for repeated indices.  Field theoretic
generalization is obvious: the discrete quantities $\{i, j, \ldots \}$ become
continuous spatial variables.  Time dependence of dynamical variables is not
explicitly indicated since all quantities are defined at the same time, but
time-differentiation is denoted by an over-dot.  Although the language of
quantum mechanics is used, with $\hbar$ scaled to unity, (``commutation,''
\etc ) ordering issues are not addressed --- so more properly speaking we are
describing a classical Hamiltonian reduction of dynamics.  Grassmann variables
are not considered, since that complication is a straightforward
generalization.  Finally total time derivative contributions to Lagrangians
are omitted whenever convenient.

Our starting point is a first-order Lagrangian formulation for the dynamics of
interest; \ie we assume that the Lagrangian is at most linear in time
derivatives.  This is to be contrasted with the usual approach, where the
starting point is a second-order Lagrangian, quadratic in time-derivatives,
and a first-order Lagrangian is viewed as ``singular'' or ``constrained.''  In
fact, just because dynamics is described by first-order differential
equations, it does not mean that there are constraints, and this is a point we
insist upon and we view the conventional position to be inappropriate.

Indeed there are many familiar and elementary dynamical systems that are
first-order, without there being any constraints:
Lagrangians for the Schr\"odinger equation and the Dirac equation are
first-order in time derivatives; in light-cone quantization, where
$x^{+} \equiv {1\over{\sqrt{2}}}(t+x)$ is the evolution coordinate, dynamics is
first-order in this ``time;'' the most compact description of chiral bosons in
two space-time dimensions is first order in time\efoot{5}.
It is clear that characterizing any of these systems as ``singular'' or
``constrained'' reflects awkward mathematics rather than physical fact.

Moreover, a conventional second order Lagrangian can be converted to
first-order form by precisely the same Legendre transform used to pass from a
Lagrangian to a Hamiltonian.  The point is that the formula
$$
\eqalignno{
H &= {\partial L\over\pdxq}\,\dot{q} - L &(1)\cr
p &\equiv {\partial L\over\pdxq}&(2)\cr
\noalign{\hbox{may also be read in the opposite direction, \hfil}}
L(p,q) &= p\,\dot{q} - H(p,q) &(3)\cr}
$$
and it is straightforward to verify that Euler-Lagrange equations for the
first-order Lagrangian $L(p,q)$ coincide with the Hamiltonian equations based
on H(p,q).
Thus given a conventional Hamiltonian description of dynamics, we can always
construct a first-order Lagrangian whose {\sl configuration space\/} coincides
with the Hamiltonian {\sl phase space\/}.

We begin therefore with a general first-order Lagrangian.
$$
L = a_i (\xi) \dxi - V(\xi)
\eqno{(4)}
$$
Note that $a_i$ has the character of a vector potential (connection) for an
Abelian gauge theory, in that modifying $a_i(\xi)$ by a total derivative
$a_i \rightarrow a_i + {\partial\over\partial\xi^i} \theta$
does not affect dynamics, since the Lagrangian changes by a total
time-derivative.  Observe further that when a Hamiltonian is defined by the
usual Legendre transform, velocities are absent from the combination
${\partial L \over \pdxi} \dxi - L$, since $L$ is first
order in $\dxi$, and $V$ may be identified with the Hamiltonian.
$$ H = {\partial L \over \pdxi} \dxi - L = V \eqno{(5)} $$
Thus the Lagrangian in (4) may be presented as
$$ L = a_i (\xi) \dxi - H(\xi) \eqno{(6)} $$
and the first term on the right side defines the ``canonical one-form''
$a_i (\xi) \, d\xi^i \equiv a(\xi)$.

To introduce our method in its simplest realization, we begin with a special
case, which in fact will be shown to be quite representative:  instead of
dealing with a general $a_i (\xi)$, we take it to be linear in $\xi^i$.
$$ a_i(\xi) = {1\over2}\xi^j\omega_{ji} \eqno{(7)} $$
The constant matrix $\omega_{ij}$ is anti-symmetric, since any symmetric part
merely contributes an irrelevant total time-derivative to $L$ and can be
dropped.  The Euler-Lagrange equation that follows from (6) and (7) is
$$ \omega_{ij} \dxj = {\partial \over \partial \xi^i} H(\xi) \eqno{(8)} $$

The development now goes to two cases.  The first case holds when the
anti-symmetric matrix $\omega_{ij}$ possesses an inverse, denoted by
$\omega^{ij}$, in which case $\omega_{ij}$ must be even-dimensional, \ie the
range $N$ of $\{i, j, \ldots \}$ is $2n=N$.  It follows from (7) that $\xi^i$
satisfies the evolution equation
$$ \dxi = \omega^{ij} {\partial \over\partial \xi^j} H(\xi) \eqno{(9)} $$
and {\sl there are no constraints\/}.
Constraints are present only in the second case, when $\omega_{ij}$ has no
inverse, and as a consequence possesses
$N'$ zero modes $z_a^i,\ a=1, \ldots, N'$.
The system is then constrained by $N'$ equations
in which no time-derivatives appear.
$$ z_a^i {\partial \over\partial \xi^i} H(\xi) = 0 \eqno{(10)} $$
On the space orthogonal to that spanned by the $\{ z_a \}$, $\omega_{ij}$
possesses an even-dimensional ($=2n$) inverse, so in this case $N=2n+N'$.

For the moment we shall assume that $\omega_{ij}$ {\sl does\/} possess an
inverse and that  there are no constraints.  The second, constrained case will
be
dealt with later.

With the linear form for $a_i(\xi)$ and in the absence of constraints all
dynamical equations are contained in (9).  Brackets are defined so as to
reproduce (9) by commutation with the Hamiltonian.
$$\eqalignno{
\dxi &= \omega^{ij} {\partial \over \partial \xi^j} H(\xi)
= i \left[ H(\xi), \xi^i \right] \cr
&= i \left[ \xi^j, \xi^i \right] \, {\partial \over\partial \xi^j} \, H(\xi)
\cr
\noalign{\hbox{This implies that we should take \hfil}}
\left[ \xi^i,\, \xi^j \right] &= i \, \omega^{ij} &\hbox{(11a)}\cr
\noalign{\hbox{or for general functions of $\xi$ \hfil}}
\left[ A(\xi), B(\xi) \right]
&= i {\partial A(\xi) \over \partial \xi^i} \, \omega^{ij} \,
{\partial B(\xi) \over \partial \xi^j} &\hbox{(11b)}\cr}$$

It is reassuring to verify that a conventional dynamical model, when presented
in the form
(3), is a special case of the present theory with $\xi^i$ comprising the
two-component
quantity $p\choose q$ and $\omega_{ij}$ the anti-symmetric $2\times 2$ matrix
$\epsilon_{ij},\ \epsilon_{12} = 1$.  Eq.~(11b) then implies $\lceil q, \, p
\rfloor = i$.

Next let us turn to the more general case with $a_i(\xi)$ an arbitrary function
of $\xi^i$,
not depending explicitly on time.  The Euler-Lagrange equation for (6) is
$$\eqalignno{
f_{ij}(\xi) \dxj &= {\partial \over \partial \xi^i} H(\xi) &{(12)}\cr
\noalign{\hbox{where \hfil}}
f_{ij}(\xi) &=
{\partial \over \partial \xi^i} a_j(\xi) -
{\partial \over \partial \xi^j} a_i(\xi) &{(13)}\cr
}$$
$f_{ij}$ behaves as a gauge invariant (Abelian) field strength (curvature)
constructed
from the gauge-variant potential (connection).
It is called the ``symplectic two-form,''
${1\over2} \, f_{ij} (\xi) \, d\xi^i \, d\xi^j$
$= f (\xi)$; evidently it is exact: $f=da$,
and therefore closed:  $df=0$.
In the non-singular, unconstrained
situation the anti-symmetric $N \times N$ matrix $f_{ij}$ has the matrix
inverse
$f^{ij}$,
hence $N=2n$, and (12) implies
$$ \dxi = f^{ij} {\partial \over \partial \xi^j} H(\xi) \eqno{(14)} $$
This evolution equation follows upon commutation with $H$ provided the basic
bracket is
taken as
$$ \left[ \xi^i,\ \xi^j \right] = i f^{ij}(\xi) \eqno{(15)} $$
The Bianchi identity satisfied by $f_{ij}$ ensures that (15) obeys the Jacobi
identity.

The result (15) and its special case (11b) can also be derived by an
alternative,
physically motivated argument.  Consider a massive particle, in any number of
dimensions,
moving in an external electromagnetic field, described by the vector potential
$a_i(\xi)$
and scalar potential $V(\xi)$.
The Lagrangian and Hamiltonian are expressions familiar from the theory of the
Lorentz force,
$$ \eqalignno{
L &= {1\over2} m \dxi \dxi + a_i(\xi) \dxi  - V(\xi) & \hbox{(16a)} \cr
H &= {1\over2m} \left( p_i - a_i(\xi) \right)^2 + V(\xi) & \hbox{(16b)}}$$
with $p_i$ conjugate to $\xi^i$.  It is seen that (4), (5) and (6) correspond
to the $m
\rightarrow 0$ limit of (16a) and (16b).  Owing to the $0 (m^{-1})$ kinetic
term in
(16b), the limit of vanishing mass can only be taken if $p_i - a_i(\xi) \equiv
m \dxi$
is constrained to vanish.  Adopting for the moment the Dirac procedure, we
recognize that
vanishing of $m \dxi$ is a second class constraint, since the constraints do
not commute,
$$\eqalignno{
\left[ m \dxi,\ m \dxj \right] &= \left[ p_i - a_i(\xi),\ p_j - a_j(\xi)
\right] \cr
&= i \, f_{ij}(\xi) \neq 0 & (17)\cr
}$$
and computing the Dirac bracket $[ \xi^i, \xi^j ]$ regains (15).

In this way we see that what one would find by following Dirac is also gotten
by our
method, but we arrive at the goal much more quickly.
Also the above discussion gives a physical setting for
Lagrangians of the form (6):  when dealing with a charged particle in an
external
magnetic field, in the strong field limit the Lorentz force term --- the
canonical
one-form --- dominates the kinetic term, which therefore may be dropped in
first
approximation.  One is then left with quantum mechanical motion where the
spatial
coordinates fail to commute by terms of order of the inverse of the magnetic
field.
More specifically, with constant magnetic field $B$ along the $z$-axis, energy
levels of
motion confined to the $x$-$y$ plane form the well-known Landau bands.  For
strong
fields, only the lowest band is relevant, and further effects of the additional
potential
$V(x,y)$ are approximately described by the ``Peierls Substitution''\efoot{6}.
This
states that the low-lying energy eigenvalues are
$$ E = {B\over2m} + \epsilon_n \eqno{(18)} $$
where ${B\over2m}$ is the energy of the lowest Landau level in the absence of
$V$, while
the $\epsilon_n$ are eigenvalues of the operator $V(x,y)$ (properly ordered!)
with
$$ i[x,\,y]={1\over B} \eqno{(19)} $$
Clearly the present considerations about quantizing first-order Lagrangians
give a new
derivation\efoot{7} of this ancient result from condensed matter
physics.\efoot{6}
[One may also verify (18) by forming $mH$ from (16b) and computing $\epsilon_n$
perturbatively in $m$.\efoot{8}]

While the development starting  with arbitrary $a_i(\xi)$ and unconstrained
dynamics
appears more general than that based on the linear, special case (7), the
latter in fact
includes the former.  This is because by using Darboux's theorem one can show
that an
arbitrary vector potential [one-form $a_i \, d\xi^i$] whose associated field
strength
[two-form $d(a_i \, d \xi^i) = {1\over2} \, f_{ij} \, d\xi^i \, d\xi^j$] is
non-singular,
in the sense that the matrix $f_{ij}$ possesses  an inverse, can be mapped by
a coordinate transformation onto (7) with
$\omega_{ij}$ non-singular.  Thus apart from a gauge term,
one can always present $a_i(\xi)$ as
$$ \eqalignno{
a_i(\xi) &= {1\over2} Q^k(\xi) \, \omega_{k\ell} \,
{\partial Q^\ell(\xi)\over\partial\xi^i} &\hbox{(20a)} \cr
\noalign{\hbox{correspondingly $f_{ij}(\xi)$ as}}
f_{ij}(\xi) &= {\partial Q^k(\xi)\over\partial\xi^i} \, \omega_{k\ell} \,
{\partial Q^\ell(\xi)\over\partial\xi^j} &\hbox{(20b)} \cr}$$
and in terms of new coordinates $Q^i$ the curvature is $\omega_{ij}$ --- a
constant and
non-singular matrix.  Moreover, by a straightforward modification of the
Gram-Schmidt argument
a basis can be constructed such that the antisymmetric $N\times N$ matrix
$\omega_{ij}$ takes the
block-off-diagonal form
$$ \omega_{ij} = \left( \matrix{0 & I \cr -I & 0 \cr} \right)_{ij} \eqno{(21)}
$$
where $I$ is the $n$-dimensional unit matrix (N=2n).
[With these procedures one can also handle the case when $a_i$ is explicitly
time-dependent --- a transformation to constant $\omega_{ij}$ can still be
made.]~
In the Appendix we present Darboux's theorem adopted for the present
application, and we
explicitly construct
the coordinate transformation $Q^i(\xi)$.
The coordinates in which the curvature two-form becomes (21) are of course the
canonical coordinates and they can be renamed $p_i,\ q^i,\ i=1,\ldots,n$.

We conclude the discussion of non-singular, first-order dynamics by recording
the
functional integral for the quantum theory.  The action of (4) obviously is
$$ I = \int a_i(\xi) \, d\xi^i - \int H(\xi) \, dt \eqno{(22)} $$
and the path integral involves, as usual, the phase exponential of the action.
The
measure however is non-minimal; the correct prescription is
$$Z = \int \Pi_i {\cal D} \xi^i \, \det{}^{1\over2} f_{jk} \, \exp i \, I\ \ .
\eqno{(23)} $$
The $\det^{1\over2} \, f_{ij}$ factor can be derived in a variety of ways:  One
may use Darboux's
theorem to map the problem onto one with constant canonical curvature (21),
where the
measure is just the Liouville measure
$\prod_{i=1}^{2n} {\cal D} \xi^i = \prod_{i=1}^{n} {\cal D} p_i \, {\cal D}
q^i$,
and the Jacobian of the transformation is seen from (20b) to be $\det^{1\over2}
f_{ij}$.  Alternatively one
may refer to our derivation based on Dirac's second class constraints,
eqs.~(16), (17), and recall that
the functional integral in the presence of second class constraints involves
the square root of the
constraints' bracket\efoot{9}.  By either argument, one arrives at (23), which
also exhibits the
essential nature of the requirement that $f_{ij}$ be a non-singular matrix.

We now turn to the second, more complicated case, where there are constraints
because $f_{ij}$ is
singular.  It is evident from the Appendix that the Darboux construction may
still be carried out for
the non-singular projection of $f_{ij}$, which is devoid of the zero-modes
(10).  This results in the
Lagrangian
$$ L = {1\over2} \, \xi^i \, \omega_{ij} \, \dxj - H(\xi,\,z) \eqno{(24)} $$
Here $\omega_{ij}$ may still be taken in the canonical form (21), but now in
the Hamiltonian there are $N'$
additional coordinates, denoted by $z_a, a=1, \ldots, N'$, arising from the
$N'$ zero modes of $f_{ij}$
and leading to $N'$ constraint equations.
$$ {\partial \over \partial z_a} \, H(\xi, z) = 0 \eqno{(25)} $$
This is the form that (10) takes in the canonical coordinates achieved by
Darboux's theorem.  The
constrained nature of the $z_a$ variables is evident:  they do not occur in the
canonical one-form
${1\over2} \, \xi^i \, \omega_{ij} \, \dxj \, dt$ and there is no
time-development for them.

In the next step, we examine the constraint equations (25) and recognize that
for the $z_a$ occurring
{\sl non-linearly} in $H(\xi, z)$ one can solve (25) for the $z_a$.  [More
precisely, this needs
$\det {\partial^2 H(\xi,z) \over \partial z_a \, \partial z_b} \neq 0.$]~
On the other hand, when $H(\xi, z)$ contains a constrained
$z_a$ variable {\sl linearly\/}, Eq.~(25) does not permit an evaluation of the
corresponding $z_a$, because (25) in that case is a relation among the $\xi^i$,
with $z_a$ absent from
the equation.
Therefore using (25), we evaluate as many $z_a$'s as possible, in terms of
$\xi^i$'s and other $z_a$'s,
and leave for further analysis the linearly
occurring $z_a$'s.  Note that this step does not affect the canonical one-form
in the Lagrangian.

Upon evaluation and elimination of as many $z_a$'s as possible, we are left
with a Lagrangian in the form
$$ L = {1\over2} \xi^i \omega_{ij} \dxj - H(\xi) - \lambda_k \Phi^k (\xi)
\eqno{(26)} $$
where the last term arises from the remaining, linearly occurring $z_a$'s, now
renamed $\lambda_k$, and
the only true constraints in the model are the $\Phi^k$, which enter multiplied
by Lagrange multipliers
$\lambda_k$.
To incorporate the constraints, it is not necessary to classify them into first
class or second class.
Rather we solve them, by satisfying the equations
$$ \Phi^k(\xi) = 0 \eqno{(27)} $$
which evidently give relations among the $\xi^i$ --- evaluating some in terms
of others.  This procedure
obviously eliminates the last term in (26) and it reduces the number of
$\xi^i$'s below the $2n$ that
are present in (26); also it replaces the diagonal canonical one-form by the
expression
$\bar{a_i} (\xi) \, d\xi^i$, where $i$ ranges over the reduced set, and
$\bar{a_i}$ is a
non-linear function obtained by inserting the solutions to (27) into (26).

The Darboux procedure must now be repeated:  the new canonical one-form
$\bar{a_i}(\xi) \, d\xi^i$, which
could be singular, is  brought again to diagonal form, possibly leading to
constraint equations, which
must be solved.  Eventually one hopes that the iterations terminate and one is
left with a completely
reduced, unconstrained and canonical system.

Of course there may be the technical obstacles to carrying out the above steps:
solving the constraints may prove too difficult, constructing the Darboux
transformation to canonical
coordinates may not be possible.  One can then revert to the Dirac method, with
its first and second
class constraints, and corresponding modifications of brackets, subsidiary
conditions on states, and
non-minimal measure factors in functional integrals.

I conclude my presentation by exhibiting our method in action for
electromagnetism coupled to matter,
which for simplicity I take to be Dirac fields $\psi$, since their Lagrangian
is already first
order.  Also I include a gauge non-invariant mass term for the photon, to
illustrate various examples of constraints.
The electromagnetic Lagrangian in first-order form reads
$$
\eqalignno{
L &= \int d\vec{r} \left\{ - \vec{E} \cdot \dot{\vec{A}}^i + i \psi^{*}
\skew{5}\dot{\psi}
- {1\over2} \left( \vec{E}^2 + \vec{B}^2  + \mu^2 \vec{A}^2 \right) \right\}
\cr
& \qquad - H_M \left( (\vnam - i \vec{A}) \psi \right) \cr
& \qquad - \int d\vec{r} \left\{ A_0 \left( \rho - \vnam \cdot \vec{E} \right)
- {\mu^2\over2} A_0^{\,2} \right\} & \hbox{(28)} \cr}
$$
Here $\vec{A}$ is the vector potential with $\vec{B} \equiv \vnam \times
\vec{A}$,
$A_0$ the scalar potential that is absent from the symplectic term, $\mu$ the
photon mass.  The matter
Hamiltonian is not specified beyond an indication that coupling to $\vec{A}$ is
through the covariant
derivative, while $\rho = \psi {}^{*} \psi$.
The Lagrangian is in the form (24); when $\mu$ is non-zero the constrained
variable $A_0$ enters
quadratically and
$$ {\delta H \over \delta A_0(\vec{r})} = 0 \eqno{(29)} $$
leads to the evaluation of $A_0$
$$ A_0 = {1\over\mu^2} \left( \rho - \vnam\cdot\vec{E} \right) \eqno{(30)} $$
so that the unconstrained Lagrangian becomes
$$
L = \int d\vec{r} \Biggl\lbrace - \vec{E} \cdot \dot{\vec{A}} + i \psi {}^{*}
\skew{5}\dot{\psi}
- {1\over2} \left( \vec{E}^2 + \vec{B}^2 + \mu^2 \vec{A}^2 + {1\over\mu^2}
\left( \rho - \vnam\cdot \vec{E} \right)^2 \right) \Biggr\rbrace
- H_M \left( \left(\vnam-i\vec{A}\right) \psi \right)
\eqno{(31)}
$$
The canonical pairs are identified as $(-\vec{E}, \vec{A})$ and $(i \psi
{}^{*}, \psi)$.
In the absence of a photon mass, the Lagrangian (28) is of the form (26), with
one Lagrange multiplier
$\lambda = A_0$.  Eq.~(29) then leads to the Gauss law constraint.
$$ \vnam\cdot\vec{E} = \rho \eqno{(32)} $$
To solve the constraint, we decompose both $\vec{E}$ and $\vec{A}$ into
transverse and longitudinal parts,
$$
\eqalignno{
\vec{E} &= \vec{E}_T + {\vnam \over \sqrt{-\nabla^2}} \, E &(33)\cr
\vec{A} &= \vec{A}_T + {\vnam \over \sqrt{-\nabla^2}} \, A &(34)\cr
\vnam\cdot\vec{E}_T &= \vnam\cdot\vec{A}_T = 0 \cr
}$$
and (32) implies $E={-1\over\sqrt{-\nabla^2}}\, \rho$. Inserting this into (28)
at $\mu^2 = 0$, we are
left with
$$\eqalignno{
L &= \int d\vec{r} \ \Biggl\lbrace - \vec{E}_T \cdot \dot{\vec{A}}_T + \rho
{1\over\sqrt{-\nabla^2}}
\skew{6}\dot{A}
+ i \psi {}^{*} \skew{5}\dot{\psi}
- {1\over2} \left( \vec{E}_T^{\,2} + \vec{B}^2 - \rho {1\over\nabla^2} \rho
\right)
\Biggr\rbrace \cr
& \qquad - H_M \left( \left( \vnam - i \vec{A}_T - i {\vnam \over
\sqrt{-\nabla^2}} A
\right) \psi \right) & \hbox{(35a)}\cr
}$$
While the constraint has been eliminated, the canonical one-form in (35a) is
not diagonal.
The Darboux transformation that is now performed replaces $\psi$ by
$\left( \exp i {1\over\sqrt{-\vec{\nabla^2}}} \, A \right) \psi$.
This has the effect of canceling $\rho {1\over\sqrt{-\nabla^2}} \,
\skew{6}\dot{A}$
against a contribution coming from $i \psi {}^{*} \skew{5}\dot{\psi}$
and eliminating $A$ from the Hamiltonian
(since $\vec{B} = \vnam \times \vec{A}_T$).  We are thus left with the
Coulomb-gauge Lagrangian
$$
L = \int d\vec{r} \, \Biggl\lbrace - \vec{E}_T \cdot \dot{\vec{A}}_T + i \psi
{}^{*} \skew{5}\dot{\psi}
- {1\over2} \left( \vec{E}_T^{\,2} + \vec{B}^2 - \rho {1\over\nabla^2} \rho
\right) \Biggr\rbrace
- H_M \left( \left( \vnam - i \vec{A}_T \right) \psi \right)
\eqno{\hbox{(35b)}} $$
without ever selecting the Coulomb gauge!
The canonical pairs are $\left(-\vec{E}_T, \vec{A}_T \right)$ and $\left( i
\psi {}^{*}, \psi \right)$.

We recall that the Dirac approach would introduce a canonical momentum $\Pi_0$
conjugate to $A_0$
and constrained to vanish.  The constraints (30) or (32) would then emerge as
secondary constraints,
which must hold so that $[H, \Pi_0]$ vanish.  Finally a distinction would be
made between the
$\mu \neq 0$ and $\mu = 0$ theories:
in the former the constraint is second class, in the latter it is first
class.\efoot{9}~ None of these
considerations are necessary for successful quantization.

Our method also quantizes very efficiently Chern-Simons theories, with or
without a conventional
kinetic term for the gauge field\efoot{10} [indeed the phase space reductive
limit of taking the
kinetic term to zero, as in (16), (17) above, can be clearly
described\efoot{11}] as well as gravity
theories in first order form, be they the Einstein model\efoot{12} or the
recently discussed gravitational gauge
theories in lower dimensions\efoot{13}.

Finally, we record a first order Lagrangian $L$ for Maxwell theory with
external, conserved sources
$(\rho, \vec{j}),~\dot{\rho} + \vnam \cdot \vec{j} = 0$,
which depends only on field strengths $(\vec{E},\,\vec{B})$ (rather than
potentials) and is self-dual in the absence of sources.
$$\eqalignno{
L &= \int \, d\vec{r} \, d\vec{r}' \,
\left( \dot{E}^i (\vec{r}) + j^i (\vec{r}) \right)
\, \omega_{ij} \, (\vec{r}-\vec{r}') \, B^j (\vec{r}') \cr
& \hbox{\qquad} - {1\over2} \int \, d\vec{r} \,
\left( \vec{E}^2 + \vec{B}^2 \right) - \int d\vec{r} \,
\Bigl( \lambda_1 (\rho - \vnam \cdot \vec{E}) + \lambda_2 \vnam \cdot\vec{B}
\Bigr)
&(36)\cr
\omega_{ij} (\vec{r}) &\equiv \epsilon^{ijk} \, {\partial_k \over \nabla^2} \,
\delta (\vec{r}) = {1\over4\pi} \epsilon^{ijk} \, {r^k \over r^3} &(37)\cr}$$
Varying the $\vec{E}$ and $\vec{B}$ fields as well as the two Lagrange
multipliers
$\lambda_{1,2}$ gives the eight Maxwell equations.
The duality transformation $\vec{E} \rightarrow \vec{B}$, $\vec{B} \rightarrow
-\vec{E}$,
supplemented by $\lambda_1 \rightarrow - \lambda_2$, $\lambda_2 \rightarrow
\lambda_1$ changes the Lagrangian by a total time derivative, when there are no
sources.  The canonical one-form is spatially non-local, owing to the presence
of
$\omega_{ij}$, which has the inverse
$$
\omega^{ij} (\vec{r}) = -\epsilon^{ijk} \, \partial_k \, \delta (\vec{r})
\eqno{(38)} $$
when restricted to transverse fields --- these are the only unconstrained
degrees of freedom
in (36).
It then follows that the non-vanishing commutator is the familiar formula.
$$
\left[ E_T^{\,i} (\vec{r}),~B_T^{\,j}(\vec{r}') \right]
= -i \, \epsilon^{ijk} \, \partial_k \, \delta(\vec{r} - \vec{r}') \eqno{(39)}
$$
This self-dual presentation of electrodynamics is similar to formulations of
self-dual fields on a line$^5$ and on a plane$^{10}$.

\vskip.5in
\centerline{\subfont Appendix}
\centerline{Darboux's Theorem}
\vskip.25in

We give a constructive derivation of Darboux's Theorem.  Specifically we show
that subject to
regularity requirements stated below, any vector potential (connection
one-form)
$a_i(\xi)$ may be presented, apart from a gauge transformation, as
$$
a_i (\xi) = {1\over2} Q^m (\xi) \, \omega_{mn} \,
{\partial Q^n (\xi) \over \partial \xi^i}
\eqno{\hbox{(A.1)}}
$$
and correspondingly the field strength $f_{ij}(\xi)$
(curvature two-form) as
$$
f_{ij}(\xi) = {\partial Q^m (\xi) \over \partial \xi^i}
\, \omega_{mn} \,
{\partial Q^n(\xi) \over \partial \xi^j}
\eqno{\hbox{(A.2)}}
$$
with $\omega_{mn}$ constant and anti-symmetric.
The proof also gives a procedure for finding $Q^m(\xi)$.
It is then evident that
a coordinate transformation from $\xi$ to $Q$ renders $f_{ij}$
constant and a further adjustment of the basis puts
$\omega_{ij}$ in the canonical form (21).

We consider a continuously evolving transformation
$Q^m(\xi; \, \tau)$, to be specified later, with the property that at $\tau=0$,
it is the
identity transformation
$$
Q^m(\xi; \, 0) = \xi^m
\eqno{\hbox{(A.3a)}}
$$
and at $\tau = 1$, it arrives at the desired $Q^m(\xi)$,
(which will be explicitly constructed).
$$
Q^m(\xi; \, 1) = Q^m(\xi)
\eqno{\hbox{(A.3b)}}
$$
$Q^m(\xi; \, \tau)$ is generated by $v^m(\xi;\,\tau)$, in
the sense that
$$
{\partial \over \partial \tau} \, Q^m (\xi; \, \tau)
= v^m \left( Q (\xi; \, \tau); \, \tau \right)
\eqno{\hbox{(A.4)}}
$$
Note, that $v^m$ depends explicitly on $\tau$.  Also we need
to define the transform by $Q^m (\xi; \, \tau)$ of quantities
relevant to the argument: connection one-form, curvature two-form
\etc~ The definition is standard: the transform, denoted by
$T_Q$, acts by
$$
\eqalignno{
T_Q \ a_i(\xi) &= a_m (Q) \, {\partial Q^m \over \partial \xi^i}
&{\hbox{(A.5a)}} \cr
T_Q \ f_{ij}(\xi) &= f_{mn} (Q) \,
{\partial Q^m \over \partial \xi^i} {\partial Q^n \over \partial \xi^j}
&{\hbox{(A.5b)}} \cr
}$$

To give the construction, we consider the given $a_i(\xi)$ to be
embedded in a one-parameter family $a_i(\xi; \, \tau)$, such
that at $\tau = 0$ we have $a_i(\xi)$ and at $\tau = 1$
we have
${1\over2} \, \xi^m \, \omega_{mi}$, where $\omega_{mi}$ is
constant and anti-symmetric.
$$ \eqalignno{
a_i(\xi; \, 0) &= a_i (\xi) &{\hbox{(A.6a)}} \cr
a_i(\xi; \, 1) &= {1\over2} \, \xi^m \, \omega_{mi}
&{\hbox{(A.6b)}} \cr } $$
It is then true that
$$ {d \over d \tau}
\left( T_Q \, a_i (\xi; \, \tau) \right)
=
T_Q \left( L_v \, a_i(\xi; \, \tau) +
{\partial \over \partial \tau} \, a_i (\xi; \, \tau) \right)
\eqno{\hbox{(A.7)}} $$
where $L_v$ is the Lie derivative, with respect to the vector
$v^m$ that generates the transformation, see (A.4).
Eq.~(A.7) is straightforwardly verified by differentiating with
respect to $\tau$, and recalling that both the transformation
and $a_i$ are $\tau$-dependent.  Next we use the identity\efoot{14}
$$ L_v \, a_i = v^n \, f_{ni} + \partial_i (v^n a_n)
\eqno{\hbox{(A.8)}} $$
and observe that when the generator is set equal to
$$ v^n (\xi; \, \tau) = - f^{ni} (\xi; \, \tau)
{\partial \over \partial \tau} \, a_i (\xi; \, \tau)
\eqno{\hbox{(A.9)}} $$
Eq.~(A.7) leaves
$$ {d\over d \tau}
\left( T_Q \, a_i \right)
= T_Q \left( \partial_i (v^n a_n) \right) \eqno{\hbox{(A.10)}}$$
Thus
${d \over d \tau} (T_Q  a_i)$
is a gauge transformation, so that $T_Q a_i$
at $\tau=0$, \ie $a_i(\xi)$, differs from its value at $\tau=1$,
\ie ${1\over2} Q^m(\xi) \, \omega_{mn} \,
{\partial Q^n(\xi)\over\partial \xi^i}$, by a gauge
transformation.  This is the desired result, and moreover
$Q^m(\xi;\,\tau)$ and $Q^m(\xi) \equiv Q^m (\xi;\,1)$
are here explicitly constructed from the algebraic definition
(A.9) for $v^n$ [once an interpolating $a_i (\xi;\,\tau)$ is chosen],
and integration of (A.4)  (the latter task need
not be easy).

Clearly (A.9) requires that $f_{ij}(\xi;\,\tau)$ possesses the
inverse $f^{ij}(\xi;\,\tau)$; hence both the starting and ending forms,
$f_{ij}(\xi)$
and $\omega_{ij}$, must be non-singular.  Also
$f_{ij}(\xi;\,\tau)$ must remain non-singular for all
intermediate $\tau$.
In fact this is not a restrictive requirement,
because one may always choose $\omega_{ij}$ to be the value of
$f_{ij}(\xi)$ at some point $\xi=\xi_0$, and then by change of
basis transform $\omega_{ij}$ to any desired form.

This description of Darboux's theorem was prepared with the
assistance of B.~Zwiebach, whom I thank.

\bigskip
\centerline{\subfont REFERENCES}
\vskip.25in
\raggedright
\parskip=\medskipamount

\item{1.}
P.~Dirac,
``Generalized Hamiltonian dynamics,''
{\sl Canad.~J.~Math.\/}~{\bf 2}, 129 (1950);
``Fixation of Coordinates in the Hamiltonian Theory of Gravitation,''
{\sl Phys.~Rev.~\/}~{\bf 114}, 924 (1959);
{\sl Lectures on Quantum Mechanics\/}
(Yeshiva University, New York, NY 1964).

\item{2.}
M.~Henneaux and C.~Teitelboim,
{\sl Quantization of Gauge Systems\/}
(Princeton University, Princeton, NJ, 1992).

\item{3.}
{\sl Constraint's Theory and Relativistic Dynamics\/},
G.~Longhi and L.~Lusanna eds. \hfil\break
(World Scientific, Singapore, 1987).

\item{4.}
L.~Faddeev and R.~Jackiw,
``Hamiltonian Reduction of Unconstrained and Constrained Systems,''
{\sl Phys.~Rev.~Lett.\/}~{\bf 60}, 1692 (1988).

\item{5.}
R.~Floreanini and R.~Jackiw,
``Self-Dual Fields as Charge Density Solitons,'' \hfil\break
{\sl Phys.~Rev.~Lett.\/}~{\bf 59}, 1873 (1987).

\item{6.}
R.~Peierls,
``Zur Theorie des Diamagnetismus von Leitungselektronen,''
{\sl Z.~Phys.\/}~{\bf 80}, 763 (1933).

\item{7.}
G.~Dunne and R.~Jackiw,
``\thinspace `Peierls Substitution' and Chern-Simons Quantum Mechanics,''
Proceedings of
the second Chia Meeting on Common Trends in
Condensed Matter and High Energy Physics,
Chia, Italy, September 1992, L.~Alvarez-Gaum\'e, A.~Devoto, S.~Fubini and
C.~Trugenberger eds. ({\sl Nucl.~Phys.~B\/}, in press); and
\hfil\break
{\sl H.~Umezawa~Festschrift\/}, S.~De Lillo, F.~Khanna, G.~Semenoff and
P.~Sodano, eds. ({\sl World Scientific\/}, Singapore, in press.)

\item{8.}
M.~Berry, private communication.

\item{9.}
P.~Senjanovic,
``Path Integral Quantization of Field Theories
with Second-Class Constraints,''
{\sl Ann.~Phys.\/}~(NY)~{\bf 100}, 227 (1976).

\item{10.}
S.~Deser and R.~Jackiw,
`` `Self-Duality' of Topologically Massive Gauge Theories,''
{\sl Phys.~Lett.\/}~{\bf 139B}, 371 (1984);
G.~Dunne, R.~Jackiw and C.~Trugenberger,
``Chern-Simons Theory in the Schr\"odinger Representation,''
{\sl Ann.~Phys.\/}~(NY)~{\bf 194}, 197 (1989).

\item{11.}
G.~Dunne, R.~Jackiw and C.~Trugenberger,
``Topological (Chern-Simons) Quantum Mechanics,''
{\sl Phys.~Rev. D\/}~{\bf 41}, 661 (1990).

\item{12.}
R.~Arnowitt, S.~Deser and C.~Misner,
``Dynamical Structure and Definition of Energy in General Relativity,''
{\sl Phys.~Rev.\/}~{\bf 116}, 1322 (1959);
L.~Faddeev,
``The energy problem in Einstein's theory of gravitation,''
{\sl Usp.~Fiz.~Nank.\/}~{\bf 136}, 435 (1982)
[Engl.~Trans.: {\sl Sov.~Phys.~Usp.\/}~{\bf 25}, 130 (1982)].

\item{13.}
R.~Jackiw, ``Gauge Theories for Gravity on a Line,''
{\sl Teor.~Mat.~Fiz.\/}~{\bf 92}, 404 (1992)
and {\sl M.~Polivanov Memorial Volume\/} (to appear).

\item{14.}
R.~Jackiw, ``Gauge Covariant Conformal Transformations,''
{\sl Phys.~Rev.~Lett.\/}~{\bf 41}, 1635 (1978).

\item{15.}
For other self-dual formulation of electromagnetism, see S.~Deser,
``Off-shell electromagnetic duality invariance,''
{\sl J.~Phys.~A\/}~{\bf 15} 1053 (1982);
J.~Schwarz and A.~Sen, ``Duality Symmetric Actions'' (preprint, April 1993).

\bye